\documentclass[prb, twocolumn, nofootinbib]{revtex4-2}

\usepackage{amssymb}
\usepackage{amsmath}
\usepackage{bbold}
\usepackage{mathrsfs} 
\usepackage{graphicx}
\usepackage{bm}
\usepackage{tabularx}
\usepackage[x11names]{xcolor}
\usepackage[unicode=true,colorlinks=true]{hyperref}
\hypersetup{urlcolor = blue, linkcolor=blue }

\usepackage[normalem]{ulem}

\begin{document}

\title{Non-local synchronization of continuous time crystals in a semiconductor}

\author{Alex~Greilich$^{1}$, Nataliia~E.~Kopteva$^{1}$, Vladimir~L.~Korenev$^{2}$, Philipp~A.~Haude$^{1}$, Linus~Kunze$^{1}$, Ben~W.~Grobecker$^{1}$, Sergiu~Anghel$^{1}$, Markus~Betz$^{1}$, Manfred~Bayer$^{1}$}

\affiliation{$^{1}$Experimentelle Physik 2, Technische Universit\"at Dortmund, Germany}
\affiliation{$^{2}$Ioffe Institute, Russian Academy of Sciences, 194021 St. Petersburg, Russia}

\date{today}

\makeatletter
\newenvironment{mywidetext}{%
  \par\ignorespaces
  \setbox\widetext@top\vbox{%
   \hb@xt@\hsize{%
    \leaders\hrule\hfil
    \vrule\@height6\p@
   }%
  }%
  \setbox\widetext@bot\hb@xt@\hsize{%
    \vrule\@depth6\p@
    \leaders\hrule\hfil
  }%
  \onecolumngrid
  \vskip10\p@
  \dimen@\ht\widetext@top\advance\dimen@\dp\widetext@top
  \cleaders\box\widetext@top\vskip\dimen@
  \vskip6\p@
  \prep@math@patch
}{%
  \par
  \vskip6\p@
  \setbox\widetext@bot\vbox{%
   \hb@xt@\hsize{\hfil\box\widetext@bot}%
  }%
  \dimen@\ht\widetext@bot\advance\dimen@\dp\widetext@bot
  \vskip\dimen@
  \vskip8.5\p@
  \twocolumngrid\global\@ignoretrue
  \@endpetrue
}%
\makeatother
\begin{abstract}

Synchronization resulting in unified collective behavior of the individual elements of a system that are weakly coupled to each other has long fascinated scientists. Examples range from the periodic oscillation of coupled pendulum clocks to the rhythmic behavior in biological systems. Here we demonstrate this effect in a solid-state platform: spatially remote, auto-oscillating electron-nuclear spin systems in a semiconductor. When two such oscillators separated by up to $40$\,$\mu$m are optically pumped, their individually different frequencies lock to a common value, revealing long-range coherent coupling. For larger separations, the synchronization breaks. The interaction distance matches the electron spin diffusion length, identifying spin transport as the coupling-mediating mechanism and establishing phase coherence over mesoscopic distances. As a consequence, a wide-area optical pump drives all oscillators within the illuminated spot into a single synchronized state, despite their inhomogeneity. This synchronization accounts for the exceptional stability of the resulting auto-oscillations, enabling collective motion in distributed spin systems and paving the way toward coherent spin networks in spintronics.

\end{abstract}

\maketitle

\section*{Introduction}

Synchronization is a universal phenomenon in which auto-oscillators adjust their rhythms to operate in unison. The defining feature is that auto-oscillators, capable of developing periodic motion, adjust their frequencies through weak interactions while retaining their individuality. Synchronization can arise mutually, when oscillators interact with each other, or be driven by an external periodic force~\cite{strogatz2003sync,Pikovsky_book}. Huygens’ 17th-century observation that two pendulum clocks on the same beam swing in phase demonstrated how mechanical systems can couple through a common medium to lock their motion~\cite{OliveiraSR2015,Willms_Huygens_2017}. Since then, synchronization has been recognized as a general principle across living and non-living systems.

In biology, it underlies fireflies flashing in concert~\cite{Buck1988}, cardiac pacemaker cells beating in harmony~\cite{TsutsuiSS2018}, and circadian clocks aligning to day-night cycles~\cite{HalbergEndo1959,AschoffScience1965}. In neuroscience, populations of neurons coordinate oscillatory firing to produce brain rhythms~\cite{ZeitlerPRE2009}. In the physical sciences, synchronization is central to micro- and nano-scale devices: spin-torque nano-oscillators can phase-lock via electrical coupling, enhancing signal power and spectral purity~\cite{Kaka2005,Mancoff2005,SinghPRA2019}. Josephson junctions~\cite{JainPhysRep1984}, laser arrays~\cite{RogisterPRL2007}, and optomechanical oscillators~\cite{BagheriPRL2013} similarly exhibit mutual entrainment, enabling coherent operation of many units. In atomic physics, synchronization occurs in driven ensembles where global coupling can overcome dephasing. As an example, Wadenpfuhl~\textit{et al.}~\cite{Wadenpfuhl_PRL_2023} showed that thermal rubidium atoms excited to Rydberg states exhibit synchronized oscillations due to mean-field interactions via a shared optical field.

Auto-oscillating systems produce narrow Fourier peaks in the frequency domain, providing an analogy to spatial crystals. So, in crystallography, sharp peaks in momentum space, known as Bragg peaks, arise from constructive interference of waves scattered by atomic planes, reflecting the underlying periodic lattice structure in space~\cite{kittel2005}. By this analogy, we refer to an auto-oscillator that exhibits spontaneous, persistent oscillations breaking the continuous time-translation symmetry as a time crystal~\cite{WilczekPRL12,sacha2020timecrystals}. A synchronized ensemble of such auto-oscillators can then also be regarded as a time crystal composed of many unit cells. These systems can be divided into discrete time crystals, which are non-autonomous systems responding at subharmonic frequencies under periodic driving~\cite{Monroe20, ZalatelRMP23}, and continuous time crystals (CTCs), which are autonomous systems sustained by constant driving and exhibiting self-oscillations. The latter have been realized in many-body systems with long-range interactions, such as Magnon Bose-Einstein condensates~\cite{AuttiPRL18}, Rubidium Bose-Einstein condensates~\cite{KongScience22}, photonic nanomaterials~\cite{Liu2023,RaskatlaRPL2024}, Rydberg gas systems~\cite{Wu_NatPhys_2024}, and polariton condensates in semiconductors~\cite{CarraroHaddad2024}.

\begin{figure*}[t!]
\begin{center}
\includegraphics[width=16cm]{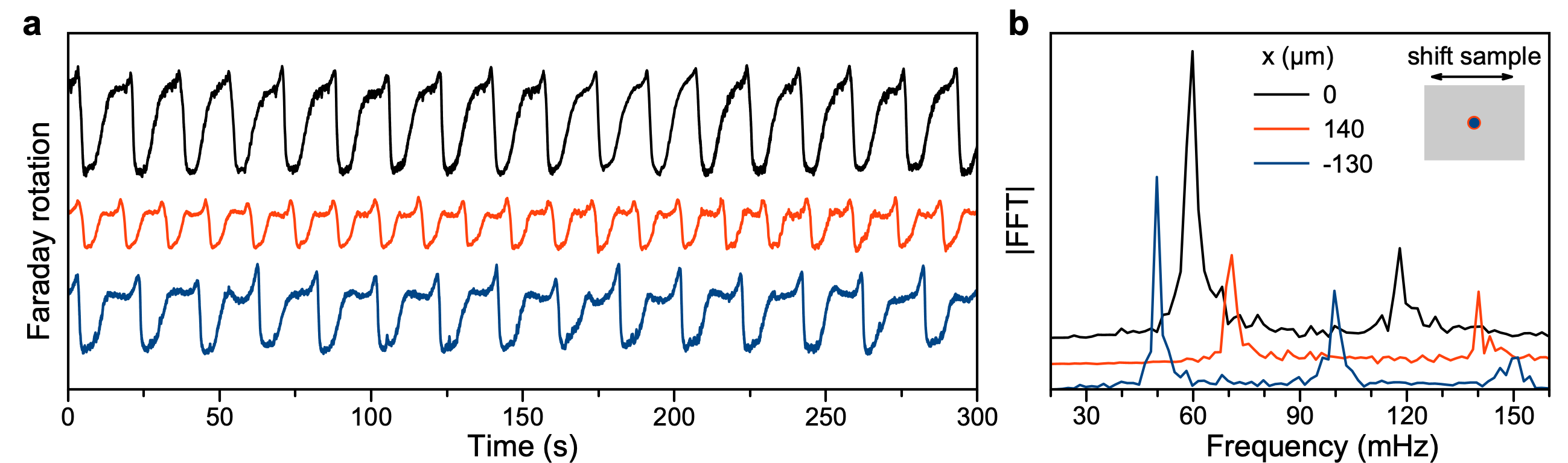}
\caption{\label{fig1} \textbf{Inhomogeneity.}
\textbf{a}, Examples of time traces measured at different sample positions. Pump and probe spots are overlapped, and the sample is shifted in the range of $-150\,\mu$m to $+150\,\mu$m along the horizontal direction ($\text{x}$). Pump spot diameter at $1/e^2$ intensity level is equal to $22\,\mu$m (FWHM of 13$\mu$m). Probe spot size at $1/e^2=12\,\mu$m (FWHM of 7$\mu$m). Pump/Probe powers are $0.15/1$\,mW, respectively.
\textbf{b}, Corresponding fast Fourier transform spectra of the traces from panel \textbf{a} measured for 10 minutes at the arbitrary position $\text{x}=0$, the top black trace; to the right side of the previous position at $\text{x}=140\,\mu$m, red trace; and to the left side at $\text{x}=-130\,\mu$m, blue trace. The difference in frequencies is related to the intrinsic sample variation.
}
\end{center}
\end{figure*}

Electron-nuclear spin systems in semiconductors can also exhibit time crystal dynamics. Recent work has demonstrated robust auto-oscillations of coupled electron and nuclear spin polarization under continuous circularly polarized excitation, a realization of a continuous time crystal~\cite{Greilich_NP24}. Furthermore, these oscillations can synchronize to an externally applied modulation when its frequency approaches their natural resonance~\cite{GreilichNC2025}, also showing a subharmonic response. Here, we demonstrate mutual synchronization between autonomous oscillators in a semiconductor system, persisting even when their natural frequencies differ by up to 40\%. Distinct spin ensembles (up to $10^9$ auto-oscillators) spontaneously lock to a common rhythm, revealing the presence of a robust collective phase. By identifying the underlying mechanism and corroborating it with theoretical modeling, we establish non-local spin synchronization as the genuine collective phenomenon. These results open new pathways toward building coherent spin networks and exploring spatially extended regimes of macroscopic synchronization.

\section*{Results}
\subsection*{Auto-oscillations}

The first step of our study is to establish robust auto-oscillations of the electron-nuclear spin system. Being an open system, the nuclei continuously dissipate energy, yet this loss can be compensated by optical pumping of the donor-bound electron spins~\cite{Greilich_NP24}. These electrons, oriented by circularly polarized pump light, transfer their spin polarization to the surrounding nuclei through the hyperfine interaction. The resulting nuclear polarization generates an Overhauser field that feeds back on the electron spins. This allows one to monitor the nuclear spin dynamics through the time evolution of the electron spin polarization, measured by the Faraday rotation of the linearly polarized probe laser. For appropriate sample design and experimental conditions, this feedback loop develops non-decaying auto-oscillations of the electron-nuclear spin system (ENSS) - a realization of the CTC~\cite{Greilich_NP24}. The main experimental parameters and sample features are detailed in the Methods and the Supplementary section~1.

Each donor electron, coupling to $\sim 10^6$ nuclear spins within its wavefunction characterized by the Bohr radius of 11\,nm~\cite{RittmannPRB2022}, defines the volume of a single auto-oscillator. In effect, the situation is better described by a flux of electrons through the donor that collectively polarize the nuclear spin ensemble, so that the experimentally accessible quantity is the average electron spin polarization at the donor. Finally, the intrinsic variation of the semiconductor sample, related to fluctuations in donor density and indium incorporation, is expected to result in variations of the observed oscillation frequencies. This feature is a significant advantage of our system due to the systematic variation of the auto-oscillation frequency across the sample.

Figure~\ref{fig1} illustrates such a variation over the sample position. Here, the pump and probe spots are tightly focused (see figure caption) and overlapped, while the sample is shifted in the horizontal direction by up to $\pm 150\,\mu$m. Figure~\ref{fig1}a demonstrates exemplary time traces for three different positions on the sample.
The fast Fourier transforms (FFT) for the 10-minute time traces are shown in Fig.~\ref{fig1}b, with the measurement positions marked in the panel. The frequency variation is pronounced. To highlight, the number of participating donors in the pump spot focused to a size of 22\,$\mu$m, is on the order of $\mathcal{O}(10^7)$. Despite that, the observed linewidth of the FFT is determined by the inverse measurement time of the non-decaying oscillations, without any dephasing that one could expect for a large number of participating oscillators. This raises the question of whether all the individual auto-oscillators, each defined by a single donor, synchronize and give rise to a macroscopic collective behavior?

\begin{figure*}[t!]
\begin{center}
\includegraphics[width=16cm]{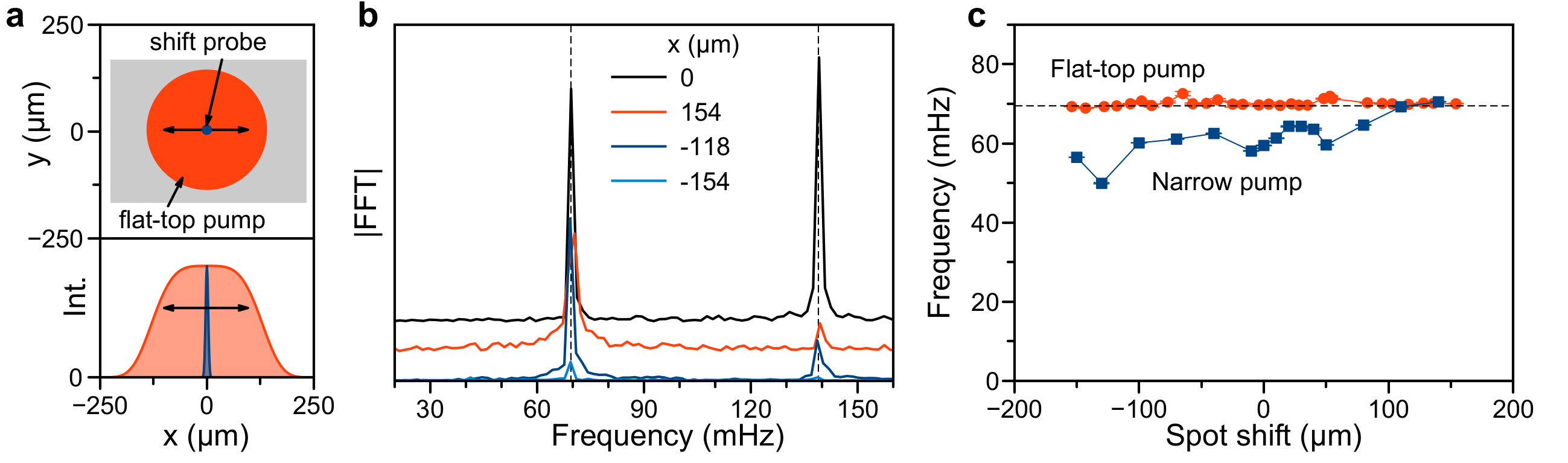}
\caption{\label{fig2} \textbf{Mutual synchronization.}
\textbf{a}, Sketch of experiment: red pump spot is prepared by a $\pi$Shaper to produce a flat-top intensity distribution on the sample, as seen in the simulated bottom part of the figure. Pump spot diameter is about 200\,$\mu$m. The Gaussian probe beam is focused down to $12\,\mu$m at $1/e^2$ (FWHM of 7$\mu$m) and can be shifted horizontally within the pump beam by the same distances as in the case of Fig.~\ref{fig1}a. The bottom part shows the simulated profiles of the lasers. Pump/Probe powers $ = 10/1$\,mW.
\textbf{b}, Examples of the fast Fourier transform spectra of the traces measured for 10 minutes at the center $\text{x}=0$, the top black trace; on the right side of the pump center at $\text{x}=154\,\mu$m, red trace; and on the left side at $\text{x}=-118\,\mu$m, blue trace, and at $\text{x}=-154\,\mu$m, light blue trace. The difference in frequencies is now negligible compared to Fig.~\ref{fig1}b, suggesting complete synchronization within the pump-spot excitation.
\textbf{c}, Summarized data for the FFT peak positions of the first harmonics for both experimental cases versus shift of the sample for tightly focused pump (blue squares) or the probe shift within a single widened pump (red circles).
}
\end{center}
\end{figure*}

\subsection*{Single-pump mutual synchronization}

To answer the question whether different auto-oscillations synchronize within a single pump spot, we designed an experiment in which we fixed the sample position and shifted the tightly focused probe spot (12\,$\mu$m) within the broadened pump (200\,$\mu$m) for the same range of sample positions as in the previous setting. The pump spot was prepared in a flat-top configuration to provide similar excitation conditions within the excitation spot diameter, as shown in Fig.~\ref{fig2}a. Figure~\ref{fig2}b again demonstrates exemplary FFTs for several probe positions inside the pump spot, showing that all observed frequencies across the pump spot are now equal. Figure~\ref{fig2}c reflects the difference in the observed FFT frequencies of the first harmonic for both cases: blue squares demonstrate the variation of the frequency over the sample measured with a tightly focused pump spot, while the red circles show that all frequencies are now equalized and are independent of the probe spot position within the single widened pump. These results can be interpreted as a clear demonstration of the mutual synchronization of spatially separated oscillators in an excited ensemble of auto-oscillators. It is essential to note that the relative deviation in frequencies synchronized within a single pump can reach up to 40\%. The number of the involved donors in that case can be estimated to be about $\mathcal{O}(10^9)$.
Importantly, the present measurements of synchronization within a single pump spot clarify the remarkable robustness of the auto-oscillations reported previously~\cite{Greilich_NP24}: their stability arises from the emergence of a many-body synchronized state, which protects the dynamics against fluctuations of external parameters.
This last demonstration is also well captured by a simplified model in which a single auto-oscillator is used to describe the whole system~\cite{DMPJETP79,Greilich_NP24, GreilichNC2025}.

\subsection*{Two-pump synchronization}

\begin{figure*}[t!]
\begin{center}
\includegraphics[width=18cm]{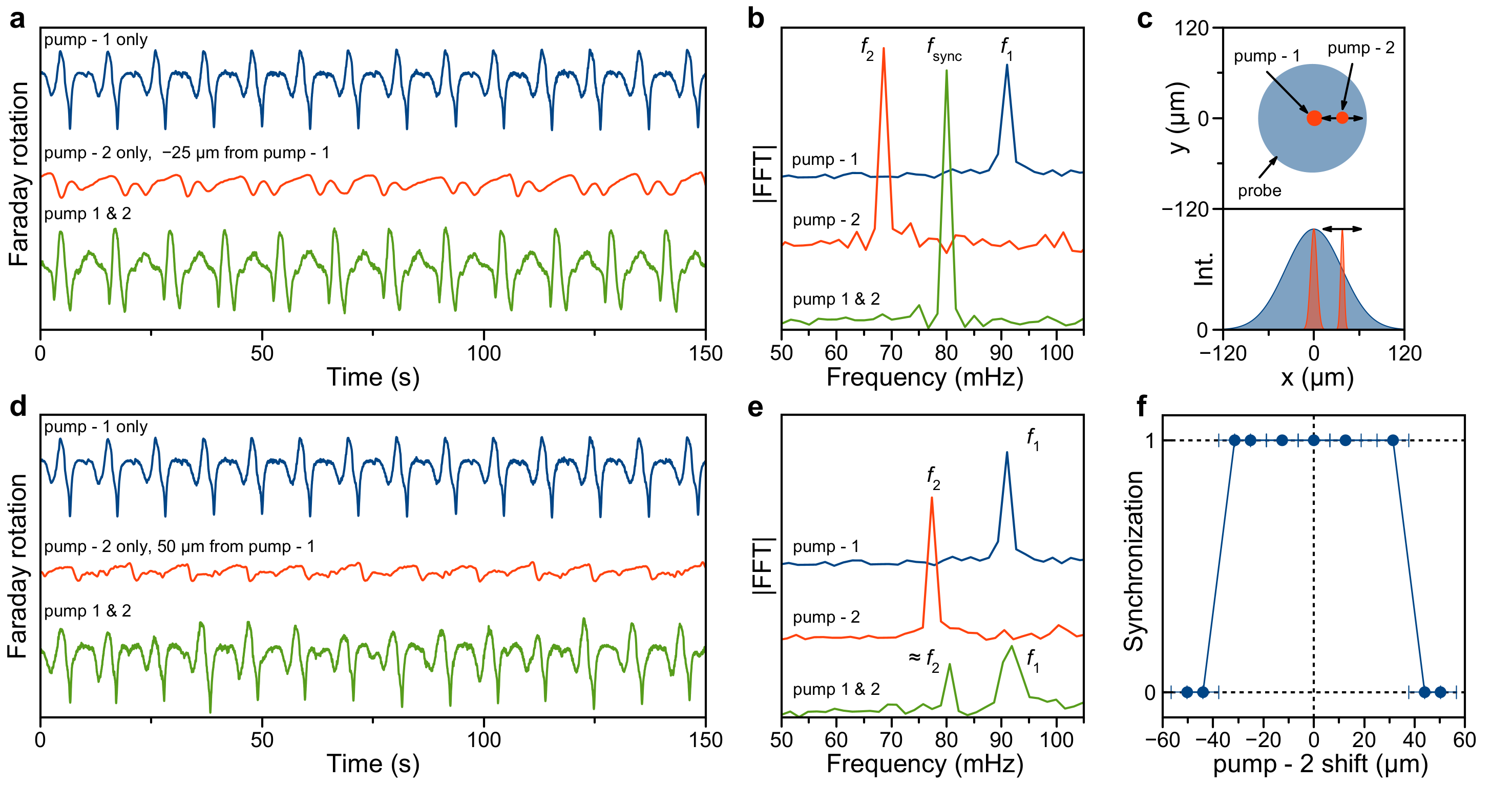}
\caption{\label{fig3} \textbf{Range of synchronization.}
\textbf{a}, Experimental time traces of auto-oscillations for three different cases: the top blue one shows the pump-1 only, the red one in the middle is the pump-2 only, shifted horizontally by $-25\,\mu$m away from the pump-1 position. Finally, the last green time trace at the bottom represents the combined application of pump-1 and pump-2. See Fig.~\ref{fig3}c for the labeling of the pumps.
\textbf{b}, First harmonic range of the fast Fourier transform spectrum of the traces presented in panel \textbf{a}, measured for 10 minutes. As one can see, each separate pump beam at its sample position induces auto-oscillations with a characteristic frequency. If both pumps are applied together, one can see a single frequency, a hallmark of synchronization. Laser powers for Pump-1/Pump-2/Pr$ = 0.1/0.1/1$\,mW.
\textbf{c}, Sketch of three Gaussian laser spots on the sample with the wide blue spot representing the probe ($1/e^2=150\,\mu$m, FWHM of 88.3$\mu$m), the central red spot is the pump-1 ($1/e^2=17\,\mu$m, FWHM of 10$\mu$m), and the red spot on the right is the pump-2 ($1/e^2=9\,\mu$m, FWHM of 5.3$\mu$m) that can be shifted horizontally. The top sketch shows a frontal view of the sample, and the bottom sketch shows the normalized intensity profile simulation of the involved beams of 38\,$\mu$m separation.
\textbf{d}, Experimental time traces of auto-oscillations for the case of the pump-2 spot shifted horizontally by 50\,$\mu$m away from the pump-1 position.
\textbf{e}, First harmonic range of the fast Fourier transform spectrum of the traces presented in panel \textbf{d}, measured for 10 minutes. Each separate pump beam induces auto-oscillations with different frequencies. When both pumps are applied together (green, bottom), two frequencies are observed at positions close to those of the single-pump beam cases, indicating no averaging and thus no synchronization.
\textbf{f}, The measured synchronization range presented for a pump-2 shift from $-50\,\mu$m to $+50\,\mu$m relative to pump-1 position. The synchronization radius is about $38\pm 3\,\mu$m measured between the pump-spot centers.
The precision of the rotation screw determines the error bars.
}
\end{center}
\end{figure*}

To investigate the processes responsible for the observed synchronization between different auto-oscillators, we modified our experimental setup by adding a second pump laser with the same photon energy as the first one. Figure~\ref{fig3}c demonstrates the setting. We use two tightly focused pump beams (shown by the two small red circles) on the surface of the semiconductor to excite separate auto-oscillations, and a wide defocused probe beam (blue circle) to read out the average electron spin polarization dynamics. The spatial position of one of the pump beams, marked as pump-2, can be precisely shifted on the sample, while pump-1 is always fixed at the center of the probe.

The blue time trace in Fig.~\ref{fig3}a demonstrates several minutes of the auto-oscillations at the pump-1 position (pump-2 is blocked), where non-decaying oscillations are observed. The corresponding FFT for a ten-minute time trace is then represented in Fig.~\ref{fig3}b by the blue curve, showing the first harmonic peak at $f_1=91\,$mHz. The red colored time trace in the middle of Fig.~\ref{fig3}a demonstrates a pump-2-only experiment, where pump-2 is, as an example, shifted by $-25\,\mu$m away from the pump-1 position, and pump-1 is blocked.
The signal intensity is shown on the same scale. It is smaller than that of the top trace of pump-1, which, in addition to the variation of the parameters of the sample, is also related to a weaker probe intensity at this spatial position, see the intensity profile variation at the bottom part of Fig.~\ref{fig3}c. Figure~\ref{fig3}b shows the corresponding FFT with $f_2=69\,$mHz, 22\,mHz lower than at the pump-1 position, by the red-colored spectrum. Finally, the green-colored trace at the bottom of Fig.~\ref{fig3}a reflects the case when both pump-1 and pump-2 are open simultaneously. As shown in Fig.~\ref{fig3}b, the corresponding FFT exhibits a single peak at $f_\text{sync}=80\,$mHz, located between $f_1$ and $f_2$, demonstrating clear synchronization. In that case, the FFT does not necessarily have to show the average frequency, but there is clearly only one peak in the 1st-harmonic range between $f_1$ and $f_2$. We emphasize that the green-colored time trace in Fig.~\ref{fig3}a represents a novel type of oscillation that cannot be composed of a combination of two contributing signals. See the Supplementary section~3 for a wider frequency range than in Fig.~\ref{fig3}a.

\subsection*{Range of synchronization}

We now continue shifting the pump beams apart. Figure~\ref{fig3}d illustrates the case where the pump-1-only signal is presented again, but the pump-2-only signal is shifted 50\,$\mu$m away from the pump-1 position (see the red-colored time trace). The green-colored bottom trace shows the signal with both pumps applied together. Figure~\ref{fig3}e shows the corresponding FFT spectra with a main difference relative to Fig.~\ref{fig3}b: the both-pumps-on case demonstrates two frequencies close to those of the positions of the contributing signals, indicating the absence of synchronization. In that example, one can argue that the green-colored trace in Fig.~\ref{fig3}d represents the superposition of both measured signals.

Finally, Fig.~\ref{fig3}f summarizes all measurements with the pump-2 position scanned across the range of $-50\,\mu$m to $+50\,\mu$m of relative shift between the pump spots. Here, the synchronization is present (1) if one observes a single frequency for the case of both pumps being applied together, and the synchronization is absent (0) if one detects both contributing single-pump frequencies under illumination with two pumps. We conclude that the synchronization range is $38\pm 3\,\mu$m taken between the central positions of the pump spots, or taking the spot sizes of the pump beams into account (given in the caption of Fig.~\ref{fig3}), the beam separation is $25\pm 3$\,$\mu$m defined as the distance between the points where the spot intensities fall to $1/e^2$ of their maximum value. In Supplementary section~3, we additionally demonstrate the appearance of intermodulation products in the superposition signal without synchronization and the FFT spectra for all pump-to-pump distances shown in Fig.~\ref{fig3}f.

\section*{Origin of synchronization}

\begin{figure*}[t!]
\begin{center}
\includegraphics[width=18cm]{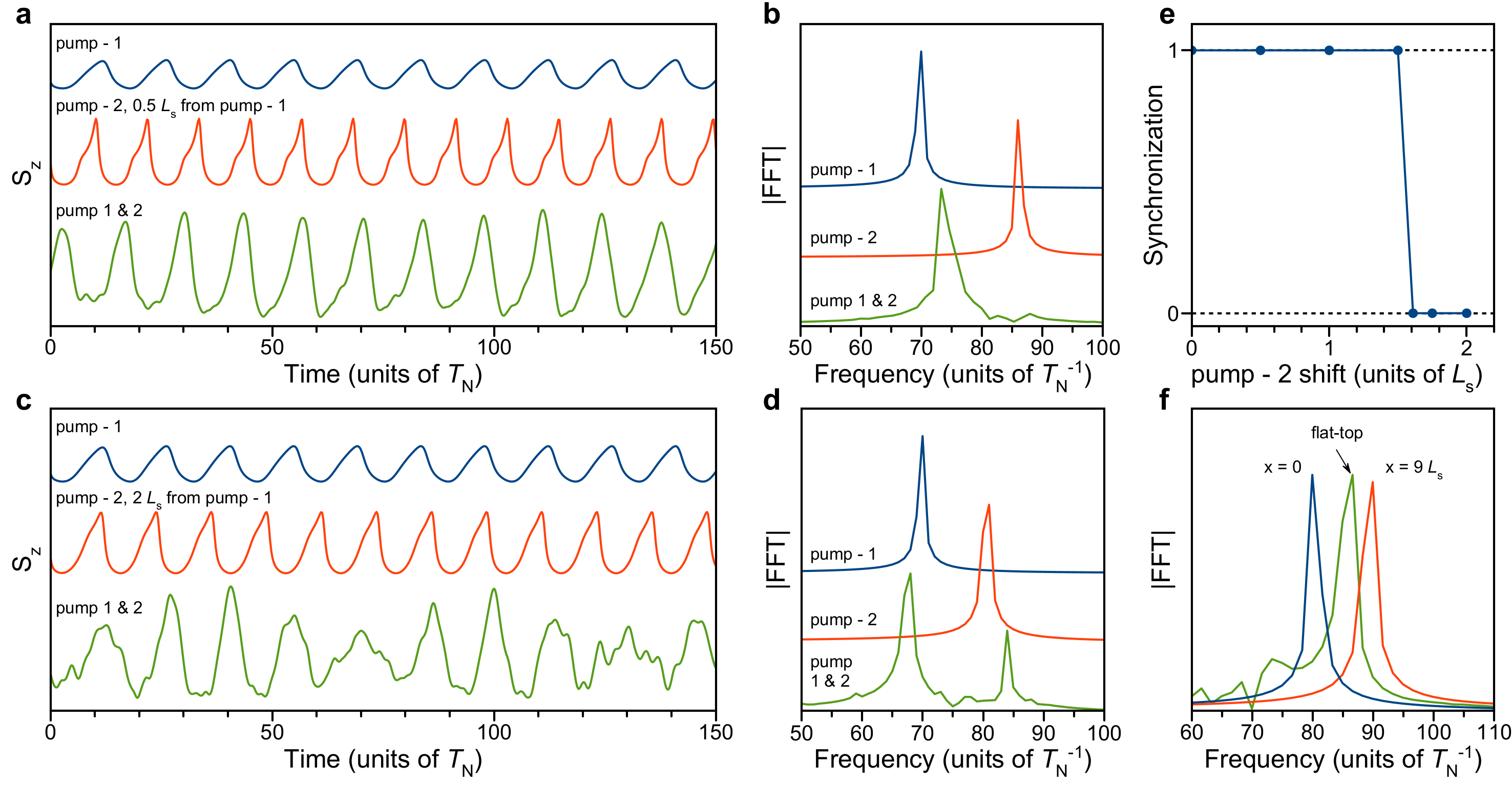}
\caption{\label{fig4} \textbf{Simulation of synchronization.}
\textbf{a}, Simulated time traces of auto-oscillations for three cases.
The top (blue) trace corresponds to pump-1 only, the middle (red) trace to pump-2 only, shifted by $0.5L_\text{s}$ from the pump-1 position.
The spatial width (FWHM) of both pump-1 and pump-2 is $0.25 L_\text{s}$.
The bottom (green) trace shows the combined excitation by pump-1 and pump-2. See Fig.~\ref{fig3}c for pump labeling. Detection is performed with a spatially wide probe covering the range $0-10 L_\text{s}$.
\textbf{b}, First harmonic of the fast Fourier transform spectra of the traces in panel \textbf{a}.
\textbf{c}, Simulated time traces of auto-oscillations for three cases: the top (blue) trace corresponds to pump-1 only, the middle (red) trace to pump-2 only, shifted by $2 L_\text{s}$ from the pump-1 position, and the bottom (green) trace to their combined excitation.
\textbf{d}, First harmonic of the FFT spectra of the traces in panel \textbf{c}.
\textbf{e}, Extracted synchronization range for pump-2 shifts between $0$ and $2 L_\text{s}$ relative to pump-1.
\textbf{f}, Synchronization in simulation under flat-top pump excitation.
The blue curve shows the first harmonic in the FFT for the spatially narrow pump-1 and narrow probe, the red curve corresponds to pump-2 and probe shifted by $9 L_\text{s}$, and the green curve shows the resulting signal under flat-top pump excitation spanning $0-10 L_\text{s}$. The detection is performed with a spatially narrow probe at $9 L_\text{s}$.}
\end{center}
\end{figure*}

For two spatially separated, auto-oscillating ENSSs, we only need a weak interaction that makes the instantaneous frequency (or phase) of one depend on the other. Let's consider the most likely mechanisms present in an $n$-doped GaAs semiconductor that could provide a connection between remote places:
\begin{itemize}

    \item {\textbf{Nuclear spin diffusion}}:
    Slow flip‑flop-driven transport of nuclear polarization spreads Overhauser fields between the spots. The diffusion constant for GaAs is tiny ($D_\text{N}\sim 10^{-13}$\,cm$^2/$s)~\cite{PagetPRB1982} and is in the same range as the one determined for our sample~\cite{RittmannPRB2022}, so the coupling is slow (seconds-minutes) but long‑lived. Using the given $D_\text{N}$ and the longest measured value of the nuclear spin relaxation time for our sample $\tau_\text{N}=210\,$s~\cite{RittmannPRB2022}, leads to a diffusion length of $L_\text{N}=\sqrt{D_\text{N} \tau_\text{N}}\approx 46$\,nm.

    \item {\textbf{Electron mediated coupling}}:
    Hopping electrons can correlate nuclear baths in separated regions, altering nuclear spin polarization and effectively coupling the ENSS phases. For a given concentration of about $10^{16}\,$cm$^{-3}$ the diffusion coefficient is estimated to be about $D_\text{hop}\approx 1$\,cm$^2$/s~\cite{Kavokin_SST2008}, which taken with the spin relaxation time of $\tau_s=120\,$ns leads to the length of $L_\text{hop}\approx 3.5\,\mu$m.

    \item {\textbf{Electron spin diffusion}}:
    Spin‑polarized free carriers created in one spot diffuse into the other, affecting the donor-bound electrons by the spin-exchange coupling~\cite{PagetPRB81}; the resulting Overhauser field feedback pulls the precession frequency and can lock the two oscillators. The spin-diffusion length in this case is given by $L_s=\sqrt{D_\text{s}\tau_\text{s}}$, where $D_s$ is the spin diffusion constant and $\tau_\text{s}$ is the electron spin relaxation time.
    Taking into account the known values of $D_\text{s}=24\,$cm$^2$/s\,~\cite{DzhioevPSS1997} and $\tau_\text{s} = 120\,$ns measured for our sample~\cite{RittmannPRB2022}, one arrives at the diffusion length $L_\text{s}\approx 17\,\mu$m.

\end{itemize}

To summarize, based on the experimentally observed synchronization range of $\sim\,40\,\mu$m, we can suggest that the primary mechanism responsible for the synchronization is the spin diffusion of free electrons~\cite{PagetPRB81, Kavokin_SST2008}.
See also Supplementary section~4 for another excluded mechanism.

To simulate it, we extend the model presented in the Refs.~\cite{DMPJETP79,DMPJETP80,Greilich_NP24}. The Bloch equation describing the spin polarization ($\mathbf{S}$) precession in the external field ($B_\text{ext}$) and the Overhauser field ($B_\text{N}$) is extended by a term accounting for spin diffusion~\cite{KKM94}:
\begin{equation}
\label{eq:ES}
\mathbf{S} = \mathbf{S}_0(x) + \frac{g\mu_\text{B}T_\text{s}}{\hbar}({\mathbf{B}_\text{ext} + \mathbf{B}_\text{N}}) \times \mathbf{S} + D_\text{s}T_\text{s}\frac{\partial^2\mathbf{S}}{\partial x^2}.
\end{equation}
Here, $D_\text{s}$ is the spin diffusion coefficient, $\mathbf{S}_0(x)$ is the average electron spin polarization in the absence of the magnetic field, created by the continuous wave pump with Gaussian or flat-top spatial profile, $\mu_\text{B}$ is the Bohr magneton, and $g$ is the electron $g$-factor. $\hbar/\mu_\text{B}gT_\text{s}$ is the half-width at half-maximum of the electron Hanle curve that is not influenced by dynamic nuclear polarization. $T_\text{s}$ is the electron spin relaxation time, $\hbar$ is the reduced Planck constant. For clarity, we restrict the diffusion to a single spatial direction ($x$).

The precession of the electron spin polarization about the total magnetic field changes the Overhauser field in time according to~\cite{DMPJETP80,OptOR_Flei_Merk}:
\begin{equation}
\label{eq:NS}
\frac{d\mathbf{B}_\text{N}}{dt} = - \frac{1}{T_\text{N}}\left(\mathbf{B}_\text{N} - \hat a\mathbf{S}\right),
\end{equation}
where $\hat a$ is the second-rank tensor describing the process of dynamic nuclear polarization. The model suggests that $\hat a\mathbf{S}$ is a linear function of $\mathbf{S}$.
The tensor has been simplified in the lowest order approximation (for details, see Methods section, Ref.~\cite{Greilich_NP24}, and the corresponding Supplementary section~5).

To account for sample variation, we introduce a spatial distribution of the hyperfine interaction parameters (see Supplementary section~5). When the system is excited by a spatially narrow Gaussian pump, auto-oscillations emerge, as shown in blue in Fig.~\ref{fig4}a, with their frequency extracted via FFT (blue curve in Fig.~\ref{fig4}b). Excitation by pump-2, positioned $0.5L_\text{s}$ away from pump-1, produces a modified oscillation pattern due to the inhomogeneous hyperfine coupling, with the corresponding frequency shift illustrated in red in Fig.~\ref{fig4}b ($L_\text{s}$ is the spin diffusion length.) When both pump-1 and pump-2 are applied simultaneously, the auto-oscillations occur at an intermediate frequency between $f_1$ and $f_2$, as seen in Fig.~\ref{fig4}b (green). This behavior represents a clear manifestation of synchronization of auto-oscillations mediated by spin diffusion.

Synchronization mediated by spin diffusion is sensitive to the spatial separation between pump-1 and pump-2. To illustrate this, pump-2 was placed $2L_\text{s}$ away from the position of pump-1. The auto-oscillations induced by pump-2 are shown in red in Fig.~\ref{fig4}c, with their corresponding frequency indicated in red in Fig.~\ref{fig4}d. When both pumps are applied simultaneously, the FFT exhibits two distinct peaks close to $f_1$ and $f_2$, as shown in Fig.~\ref{fig4}d, with the corresponding time trace also presented in Fig.~\ref{fig4}c. The emergence of two distinct frequencies under simultaneous excitation by pump-1 and pump-2 is a clear indication of the loss of synchronization. Figure~\ref{fig4}e shows the synchronization range versus pump-2 shift. Synchronization persists up to $1.5L_\text{s}$, consistent with experiment (Fig.~\ref{fig3}f). The simulations reveal that the difference between the maximum and minimum values of the hyperfine coupling parameter significantly affects the synchronization threshold (see Supplementary section 5). So, for more similar initial frequencies, the synchronization range expands.

For completeness, we also simulated synchronization under a flat-top pump covering the spatial range from $x=0$ to $x=10 L_\text{s}$. To illustrate the spatial variation of the hyperfine interaction, we show the auto-oscillation frequencies $f_1$ and $f_2$ obtained for a spatially narrow pump placed at $x=0$ and $x=9L_\text{s}$, respectively (blue and red curves in Fig.~\ref{fig4}f). Under flat-top excitation, the auto-oscillation frequency locks to an intermediate value between $f_1$ and $f_2$ across the entire range $0 \leq x \leq 10 L_\text{s}$ (see Fig.~\ref{fig4}f)). Flat-top excitation demonstrates that spin diffusion promotes global synchronization across an inhomogeneous sample, consistent with the experimental observations.

Together with the agreement between experiment and simulation, the results of simulations establish electron spin diffusion as the microscopic mechanism underlying the collective synchronization of auto-oscillations. However, to obtain a more complete picture, we also determined the diffusion length using the independent method of time-resolved magneto-optical Kerr rotation microscopy (see the Method section and the Supplementary section~6). The obtained spin diffusion value of $D_\text{s} = 28\pm 1\,$cm$^2$/s is very close to the previously reported one $D_\text{s}=24\,$cm$^2$/s\,~\cite{DzhioevPSS1997} cited above. This corresponds to a diffusion length of $L_\text{s} = 18.3 \pm 0.3\,\mu$m in our sample, confirming the observations.

\section*{Conclusion}

Our results establish that spatially separated electron–nuclear spin systems in a semiconductor can mutually synchronize through electron spin diffusion, locking their oscillations over mesoscopic distances. This collective coherence persists under wide-area excitation, where an extended ensemble of auto-oscillators contributes to a single synchronized state despite the environmental variation. Importantly, the emergence of such macroscopic synchronization is itself the key factor that stabilizes the auto-oscillations, protecting them against fluctuations and disorder that would otherwise disrupt coherent dynamics. Demonstrating that diffusive spin currents can mediate both long-range phase coherence and enhanced stability highlights a new route for engineering robust networks of coupled spin oscillators in solid-state platforms. Furthermore, integrating electrostatic gating or lateral drift control would convert this passive non-local link into an actively tunable coupling channel. Such controllable networks may serve as a basis for analog neuromorphic architectures, where information is processed through collective phase dynamics rather than discrete logic states. Beyond deepening our understanding of non-equilibrium collective dynamics, spin-mediated synchronization offers prospects for coherent spintronic architectures and on-chip time-crystal-based devices.


%

\section*{Methods}
\subsection*{Setup}
Ref.~\cite{Greilich_NP24} and the corresponding Supplementary sections~1 and 2 describe our sample and experimental setup in all details. Here, we reiterate the most relevant and deviating parts. We use the continuous wave laser diode emitting at 1.579\,eV (785\,nm) for the pump laser, which is then routed through a quarter-wave plate to create circular polarization. The linearly polarized probe laser is produced by the continuous wave Ti:Sapphire ring-laser and is fixed at the 1.426\,eV (869.4\,nm). Both lasers are combined on a non-polarizing beam splitter and are focused by a single lens at the sample.
Depending on the experiment, we also split a portion of the pump laser to produce an additional pump beam, which hits the sample at an angle of approximately 30 degrees. For the two-pump version presented in Fig.~\ref{fig3}, we use an additional beam as a pump-1, which is fixed at the center of the probe beam at the sample. The second part of the pump beam, which is routed through the rotatable NPBS and combined with the probe, is then referred to as a pump-2. The NPBS controls the pump-2 position at the sample by the tip-tilt-rotation stage (TTR001 Thorlabs). For the experiment, presented in Fig.~\ref{fig1}, we use only one pump beam and shift the whole cryostat by the micrometer stages. For the experiment in Fig.~\ref{fig2}, the pump-1 beam is going through the $\pi$Shaper (AdlOptica Focal-$\pi$Shaper), which converts the Gaussian beam to a flat-top profile. The position of that beam is fixed at the sample, while the probe position is shifted within the pump.

The sample is mounted in a helium flow cryostat at $T=6\,K$ in the center of two orthogonal pairs of electromagnet coils, generating the magnetic field components $B_x = -1$\,mT and $B_z = 0.176\,$mT.

The GaAs substrate of the sample completely absorbs the pump laser, while the probe laser is transmitted and then analyzed by the polarization bridge, which is built with a half-wave plate and a Wollaston prism. A balanced photodiode is used to measure the Faraday rotation of the plane of the linearly polarized probe.

We have used the Beam-master scanning multiple knife-edge device (Coherent) to characterize the beam's spot sizes at the sample position. It allows for measuring beam spot diameters as small as 3\,$\mu$m with 0.1\,$\mu$m resolution and larger beams up to 3\,mm with 1\,$\mu$m resolution.

More details and schematics are provided in the Supplementary section~2.

\subsection*{TR-MOKE}
For the spatial mapping and extraction of the spin diffusion coefficient, we use the following experimental configuration. Initial laser pulses with a temporal width of $\sim 35$\,fs are derived from a 60-MHz mode-locked Ti-sapphire oscillator. They are split into pump and probe paths and tuned by grating-based pulse shapers, resulting in pulses with a bandwidth of $\sim 0.5$\,nm and temporal resolution of $\sim 1$\,ps. Pump pulses are modulated between left and right circular polarizations by an electro-optic modulator. The linearly polarized probe pulses are collinear with the pump and focused on the sample surface through a microscope objective. The FWHM diameters of pump and probe pulses are $3\pm 0.1$ and $1\pm 0.1$\,$\mu$m, respectively. Kerr rotation is measured using balanced photodiodes connected to a lock-in amplifier, referenced to the modulation frequency. The probe polarization is resolved using a half-wave plate and a Wollaston prism. A mechanical delay stage adjusts the delay time between the pump and probe. The spatial overlap of the pump with the fixed and centered probe is adjusted through a lateral translation of the input lens of the beam-expanding telescope in the pump path~\cite{Henn2013UltrafastSupercontinuum,Henn2014TimeSpatiallyResolved}. All measurements are performed with the pump photon energy set to $E_\text{pu} = 1.459$\,eV and the peak power density of $4.7$\,MW/cm$^2$. The probe photon energy is $E_\text{pr} = 1.43$\,eV with peak power density of $1.2$\,MW/cm$^2$.

\subsection*{Details on simulation model}
The quadrupole unperturbed nuclear spin sublevels create the Overhauser field $\mathbf{B}_\text{N}^0$ as in pure, unstressed GaAs. $\mathbf{B}_\text{N}^0$ is aligned along the external magnetic field and compensates for the Zeeman splitting of the electrons in the external magnetic field. Due to the strong deformation caused by the indium incorporation, the spin of the $i$-th nucleus is oriented along the main local axis $\mathbf{n}_i$ of the tensor describing the quadrupole interaction rather than along the external magnetic field. The contribution of these nuclei to the total Overhauser field is $\textbf{B}_\text{Q} = \sum_i a_i(\textbf{S}\textbf{n}_i)\textbf{n}_i$, where the summation is carried out over all quadrupole perturbed nuclei within the electron localization volume around a donor. For an isotropic distribution of the axes, the field can be written as $\textbf{B}_\text{Q} = a_\text{N}\textbf{S}$. Therefore,
$\hat{a}$ can be reduced to the simplified form: $\hat{a}\textbf{S} = \textbf{B}_\text{Q} + \textbf{B}_\text{N}^0 = a_\text{N}\textbf{S} + b_\text{N}(\textbf{S}\textbf{h})\textbf{h}$, where $b_\textbf{N}$ is the parameter of the hyperfine interaction between the electrons and the nuclei, and $\textbf{h}$ is the unit vector of the externally applied magnetic field. The tensor components are: $\hat{\alpha}_{\alpha\beta}=a_\text{N}\delta_{\alpha\beta}+b_\text{N}\text{h}_\alpha \text{h}_\beta$, with $\alpha, \beta = x,y,z$ coordinates.

The solution is integrated over a time window of $\sim3T_\text{s}$, since the nuclear spin system has the relaxation time of $T_\text{N} \gg T_\text{s}$ and therefore experiences a stationary electron-nuclear polarization. This approach avoids numerical difficulties associated with solving the stationary Bloch equation directly.

The simulation algorithm consisted of three steps. In the first step, the electron spin polarization and its diffusion were calculated in the presence of the external magnetic field only. The stationary Bloch equation given by Eq.~\eqref{eq:ES} leads to a divergent solution.
To eliminate this divergence, the time-dependent Bloch equation was solved, and the resulting steady-state value was obtained by averaging over a time interval of $3T_s$.
This procedure is justified because the nuclear spin relaxation time is much longer than the electron spin lifetime. The resulting spatial distribution of electron polarization defined the spatial profile of the Overhauser field, which exhibited auto-oscillations according to Eq.~\eqref{eq:NS}. In the final step, the diffusion of the electron spin polarization was computed in the total magnetic field (Eq.~\eqref{eq:ES}), given by the sum of the external field and the Overhauser field obtained in the previous step.

All calculations were performed numerically in Matlab using standard built-in libraries.
The Overhauser field was calculated using the ode23 solver, while spin diffusion was simulated with the pdepe solver on a $600\times4000$ grid (space $\times$ time),
covering a spatial range of $11.5L_\text{s}$ and a temporal range of $1000T_\text{N}$ with Neumann boundary conditions.
The resulting signal included a transient region corresponding to the stabilization of auto-oscillations,
containing approximately 250 points in the time series.
The Fourier spectra presented in the main text were calculated after removing these transient points.

\section*{Acknowledgments}
A.G. and M.B. acknowledge support by the BMBF project QR.N (Contract No.16KIS2201). The Resource Center "Nanophotonics" of Saint-Petersburg State University provided the epilayer sample.
We acknowledge the use of ChatGPT (version 5.0, OpenAI) to assist in proofreading and stylistic polishing of the manuscript. The authors take full responsibility for the content of this publication.

\section*{Author contributions}
A.G. and N.E.K. contributed equally to this paper. A.G. built the experimental apparatus and performed the measurements together with Ph.A.H. and L.K. B.W.G. and S.A. provided the measurements and analysis of the scanning Kerr microscopy. N.E.K. and A.G. analyzed the data. N.E.K. and V.L.K. provided the theoretical description. A.G., N.E.K., and V.L.K. wrote the manuscript in close consultation with M.~Betz and M.~Bayer.

\section*{Competing interests}
The authors declare no competing interests.

\clearpage
\newpage

\setcounter{figure}{0}
\setcounter{table}{0}
\setcounter{page}{1}
\setcounter{section}{0}
\makeatletter
\renewcommand{\thepage}{S\arabic{page}}
\renewcommand{\theequation}{S\arabic{equation}}
\renewcommand{\thefigure}{S\arabic{figure}}
\renewcommand{\thesection}{S\arabic{section}}
\renewcommand{\bibnumfmt}[1]{[S#1]}
\renewcommand{\citenumfont}[1]{S#1}


\begin{titlepage}
    \centering
    {\textbf{\large{Supplemental Materials for}}\par}
    {\textbf{\large{Mutual synchronization of continuous time crystals in a semiconductor}}\par}
    \vspace{1.5cm}
    \vfill
\end{titlepage}

\maketitle

\section{Sample characterization}

\begin{figure*}[h]
\begin{center}
\includegraphics[width=15cm]{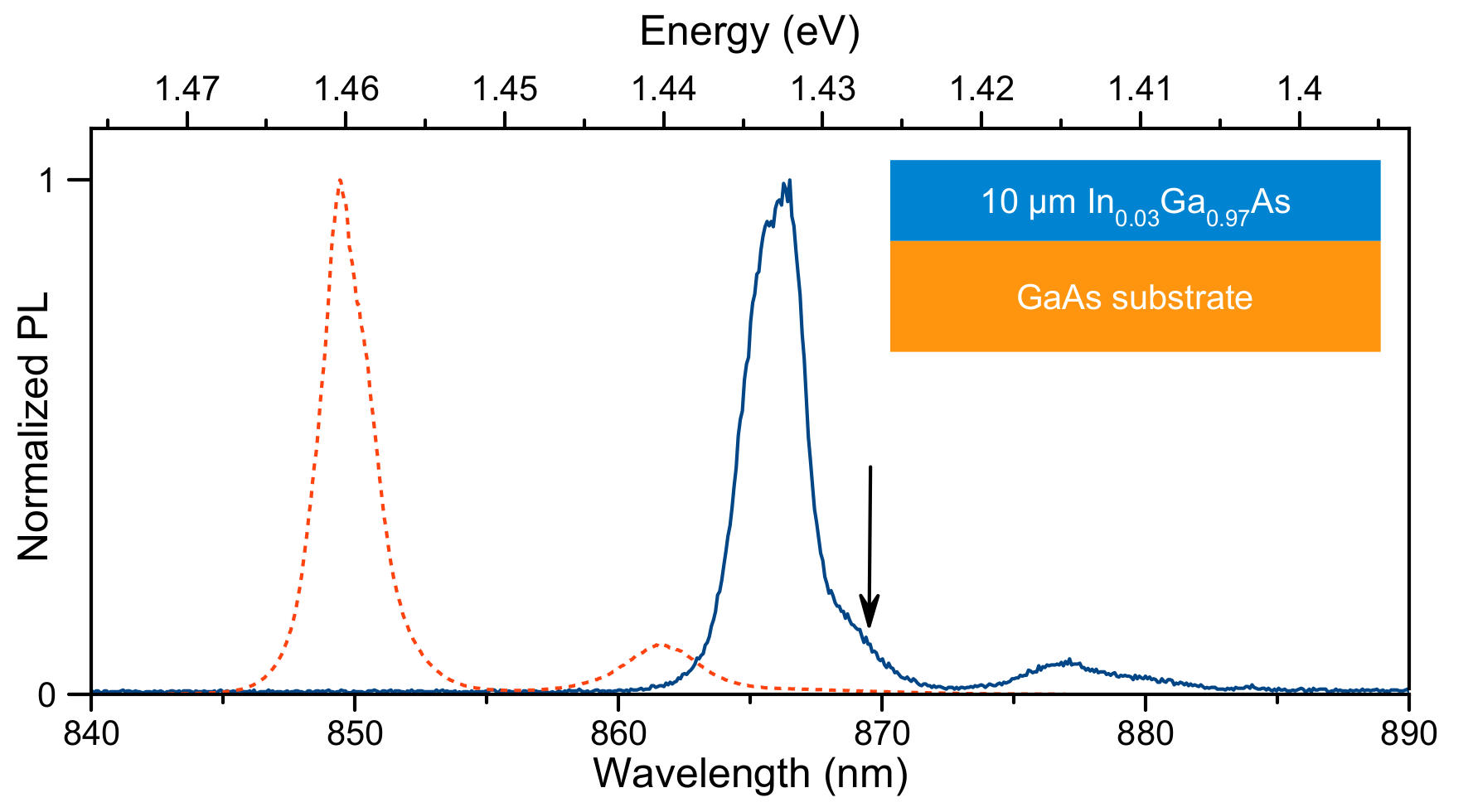}
\caption{\label{figSI0} \textbf{PL and Sample.}
Normalized PL of two samples. The red dashed line shows the PL of the sample on the thick GaAs substrate. The blue solid line shows the sample with the reduced thickness of the GaAs substrate. The black arrow shows the position of the probe wavelength. Pump is at $E_\text{exc}=1.579\,$eV (785\,nm), and the probe $E_\text{pr}=1.426\,$eV (869.4\,nm). The insert shows the sample structure with 10\,$\mu$m epilayer and GaAs substrate.}
\end{center}
\end{figure*}

Figure~\ref{figSI0} shows the normalized photoluminescence (PL) of the used sample. In comparison to previous publications~\cite{Greilich_NP24, GreilichNC2025}, we have used the same sample but etched a majority of the GaAs substrate away to test its effect on the behavior of the auto-oscillations. It has caused a shift in the PL emission to a higher wavelength. The dashed red line demonstrates the position of the PL in its initial form, while the blue solid line shows the position after the etching process. It had no observable effect on the doping and overall CTC stability and operation, except that the main frequency of the CTC (and corresponding higher harmonics) has been decreased, from $\sim 160$\,mHz to $\sim 90$\,mHz.

\section{Experimental setups}

\begin{figure*}[h]
\begin{center}
\includegraphics[width=11cm]{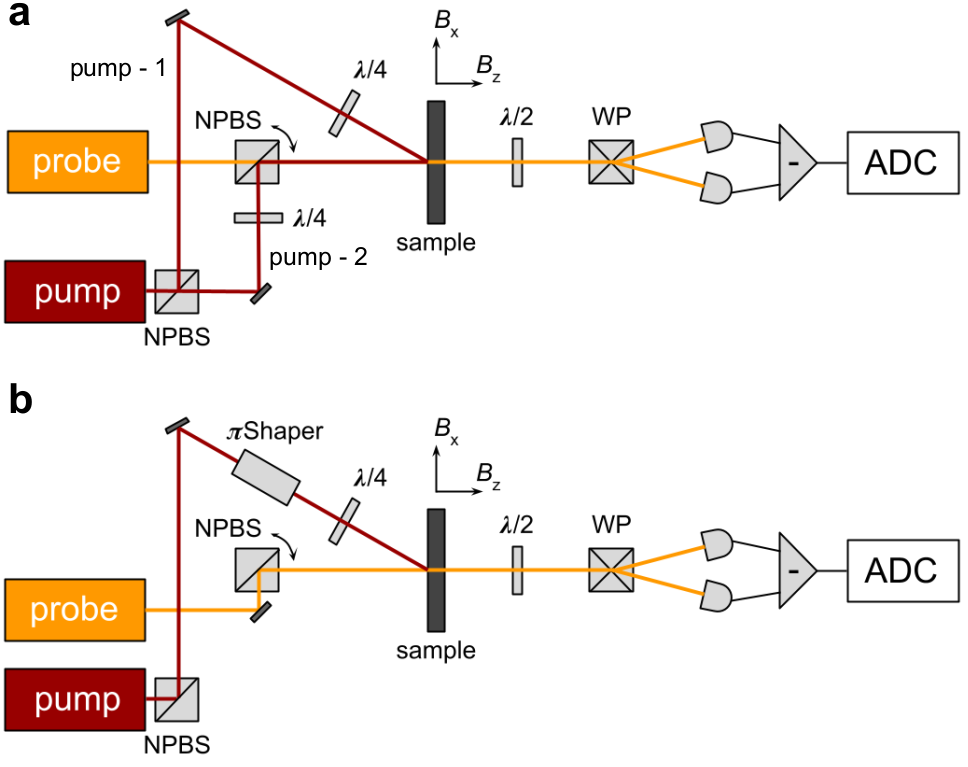}
\caption{\label{figSI1} \textbf{Schematics of the experimental setups.}
\textbf{a}, The setup for the two-pump configuration. The additional pump beam (pump-1, coming under an angle) is blocked for the measurements presented in the main text Fig.~1, and the $\mu$m-stages move the cryostat. For the two-pump experiments presented in Fig.~3 of the main text, pump-1 is focused on the center of the probe beam while pump-2 can be shifted at the sample. \textbf{b}, Configuration for a flat-top pump measurements, presented in main text Fig.~2. The elements are: a non-polarizing beam splitter (NPBS), $\lambda/2$ and $\lambda/4$ are the half- and quarter-wave plates, a Wollaston prism (WP), and a $\pi$Shaper to convert the Gaussian profile to a flat-top beam profile. The rotation of the NPBS in front of the sample was used to shift the spot reflected on the NPBS onto the sample. The energy of the pump laser is $E_\text{pu}=1.579\,$eV (785\,nm), and of the probe $E_\text{pr}=1.426\,$eV (869.4\,nm).}
\end{center}
\end{figure*}

The experimental setup for a two-pump configuration, presented in Fig.\,3 of the main text, is shown in Fig.~\ref{figSI1}a.

We use the continuous wave laser diode emitting at 1.579\,eV (785\,nm) for the pump laser, which is then routed through the laser power stabilizer (NEL03A Thorlabs). We split the pump beam into pump-1 and pump-2 using NPBS. For the pump-1 beam, we use a lens with $f=35$\,mm, a $20\,\mu$m pinhole, and collimate the beam with a 150\,mm lens. After that, we have beam routing to $\lambda/2 - $Glan-Taylor (GT)$ - \lambda/4$ with wide apertures (15\,mm) and focus it at the sample using an achromatic lens with a focal length of $f=100$\,mm. The beam is directed at an angle of approximately $30^\circ$ and is focused to a diameter of $17\,\mu$m at a $1/e^2$ intensity roll-off. The FWHM is the diameter at $1/e^2$ times $\sqrt{\ln 2}/\sqrt{2}$.

The pump-2 is prepared by using $35\,$mm lens, then $20\,\mu$m pinhole, and 150\,mm collimating lens. It is then going through GT$ - \lambda/4$ with wide apertures (pupil diameter 15\,mm) to the NPBS, where it is combined with the probe beam path.

The linearly polarized probe laser is created by the continuous wave Ti:Sapphire ring-laser and is fixed at the energy of 1.426\,eV (869.4\,nm).

The pump-2 and probe beams are focused by a single achromatic lens of 60\,mm focal length at the sample.

The 10\,$\mu$m epilayer and the remaining GaAs substrate of the sample absorb entirely the pump laser. In contrast, the probe laser is transmitted and then analyzed by the polarization bridge, which is built with a half-wave plate and a Wollaston prism. A balanced photodiode is used to measure the Faraday rotation of the plane of the linearly polarized probe, see Fig.~\ref{figSI1}a. The differential signal is digitized with 4\,ms time steps and stored on the PC. All presented time series in the main text were smoothed using a 20-point moving average.

For the measurements of sample inhomogeneity (see Fig.~1), we block the pump-1 beam (keep the pump-2 beam on) and shift the cryostat using the $\mu$m stages.

For the flat-top pump experiment (see Fig.~2), we use the scheme shown in Fig.~\ref{figSI1}b. Here, the pump-2 beam is blocked, while the probe is routed in the side entrance of the NPBS to be shifted at the sample. Pump-1 is transformed using the $\pi$Shaper to produce the flat-top beam at the sample.

To recalculate the shift of the pump by rotating the stage with NPBS in Fig.~\ref{figSI1}a, one has to consider the following: a shift at the sample $x=f\cdot2\cdot \tan(\theta)$, and for small $\theta$, $\tan(\theta)\approx \theta$. $f$ - is the focal length of the lens, in our case 60\,mm. One full rotation of the $\mu$m-screw at the rotation stage corresponds to 500\,$\mu$m scale reading and 1.5$^\circ$ of rotation about vertical axis, or $1\,\mu$m$ = 3\times 10^{-3}$\,deg. Therefore, a shift at the sample $x=f\cdot3\times 10^{-3}\cdot 2 \pi / 180\cdot s$[$\mu$m], with $s$ - being the micrometer scale of the rotation screw. With $f=60\,$mm it is: $x=2\pi\cdot s$[$\mu$m]. The scaling factor was additionally confirmed by placing the $\mu$m scale ruler at the sample position and reading the spot shift using the camera.

As the probe beam also passes through the same (rotating) NPBS, it is expected to exhibit a weak horizontal shift with NPBS rotation, which is an order of magnitude smaller than that of the reflected pump beam and is therefore irrelevant in all experiments.

\section{Extended Data}

\begin{figure*}[h]
\begin{center}
\includegraphics[width=18cm]{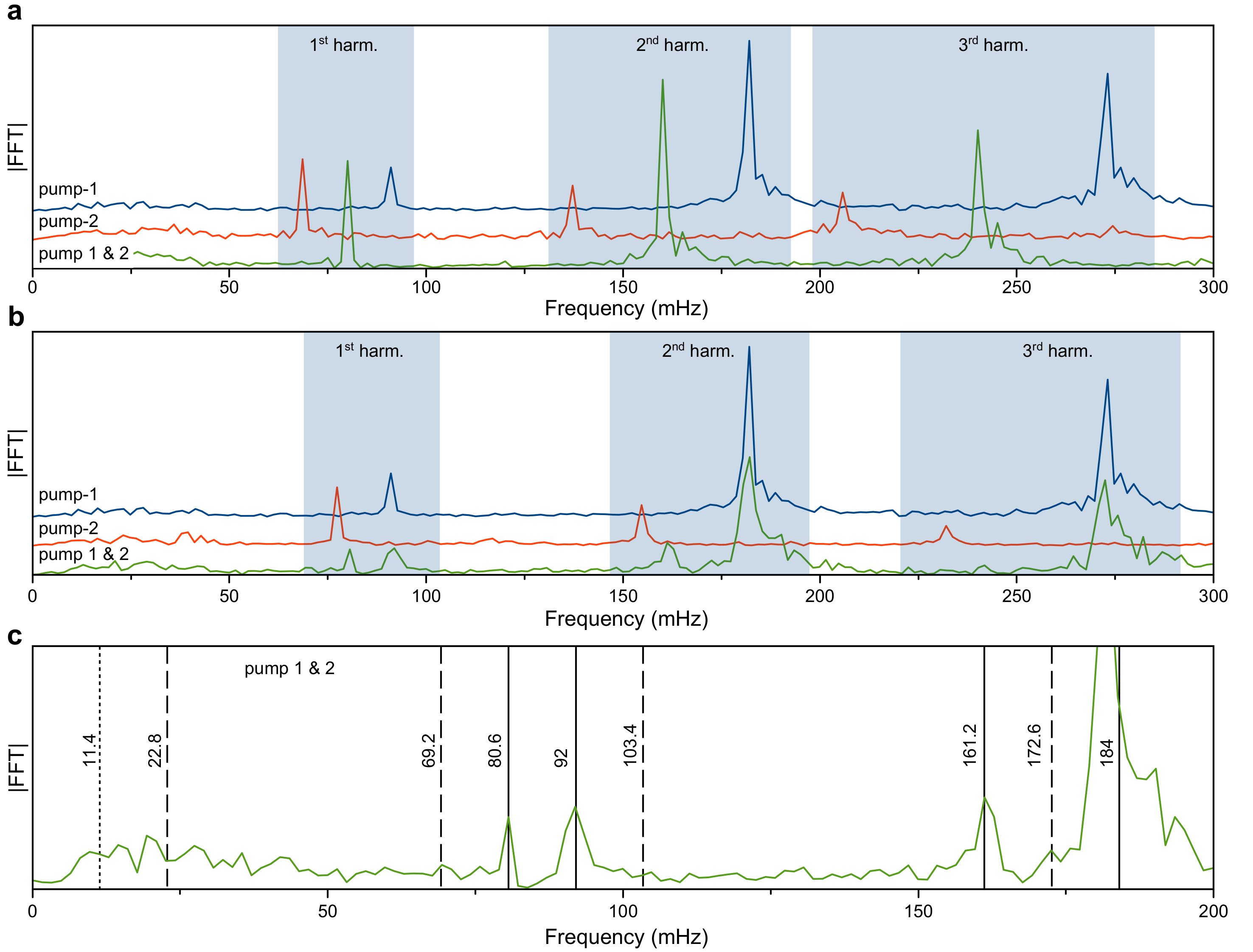}
\caption{\label{figSI2} \textbf{Extended FFT data.}
\textbf{a}, Extended frequency range of the Fig.~3b demonstrating the range of the first three harmonics in the case of synchronization (1:1 synchronization).
\textbf{b}, Extended frequency range of the Fig.~3e demonstrating the range of the first three harmonics in the absence of synchronization (free-running).
\textbf{c}, Zoomed-in FFT for the case without synchronization, with the solid lines marking the main frequencies ($f_1, f_2, 2f_1, 2f_2$), a short-dashed line marking the beat (difference), and long-dashed lines marking the intermodulation products with $n,m\le 2$.
}
\end{center}
\end{figure*}

\begin{figure*}[h]
\begin{center}
\includegraphics[width=14cm]{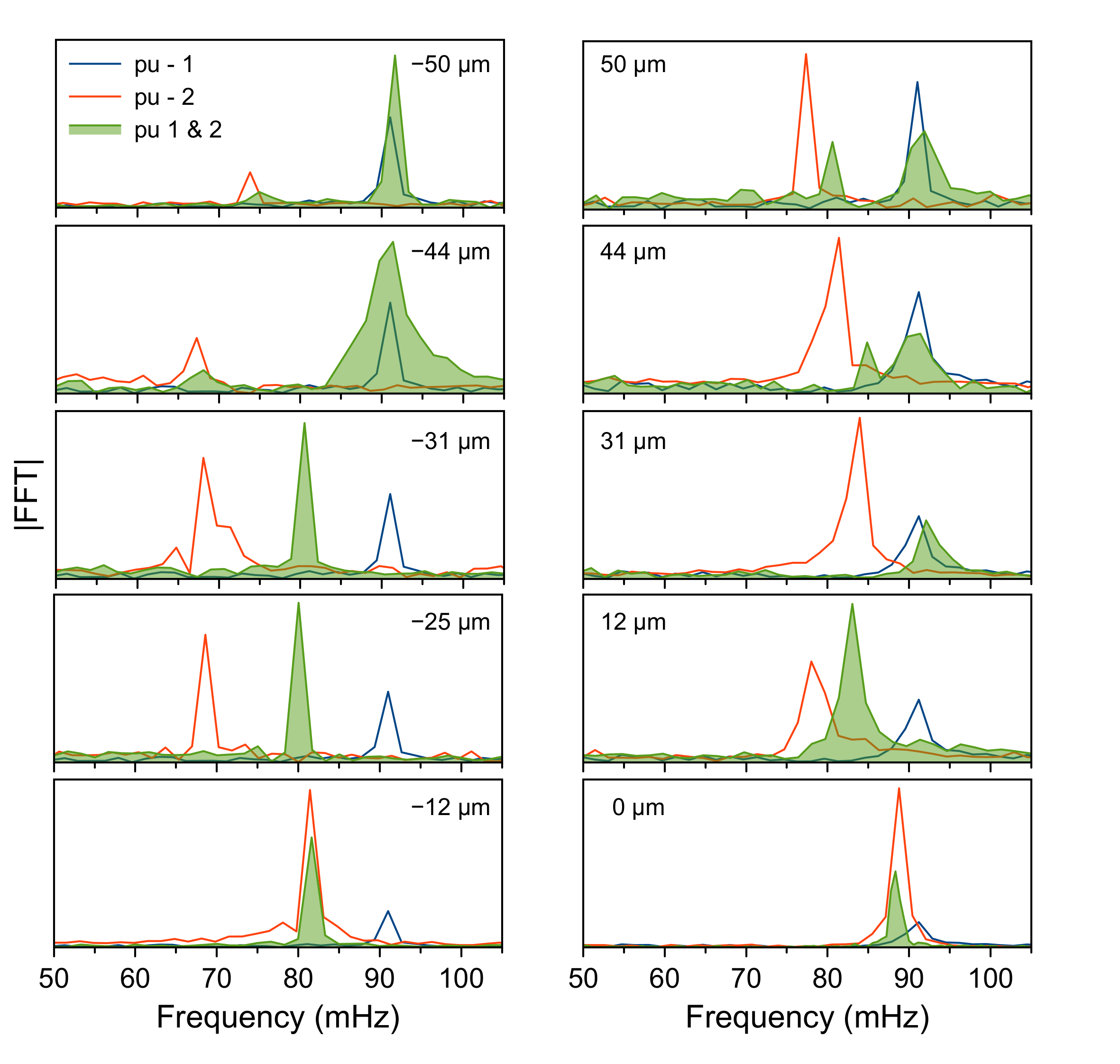}
\caption{\label{figSX} \textbf{FFT at different pump separations.}
FFT spectra for pump-1-only (blue), pump-2-only (red), and both pumps together (green shaded) cases for different pump-spot separations.
}
\end{center}
\end{figure*}

Depending on the symmetry and strength of coupling, two auto-oscillators can produce the following response in the FFT:

\begin{itemize}
\item \textbf{1:1 synchronization:} a single, narrow line at $f_{\mathrm{sync}}$; harmonics at integer multiples of $f_{\mathrm{sync}}$; intermodulation sidebands largely vanish, see Fig.~\ref{figSI2}a.
The synchronization frequency between two auto-oscillators is determined by the balance of coupling strengths~\cite{Kuramoto_1975}, nonlinear frequency shifts, dissipation, and nonreciprocal feedback~\cite{Fruchart_Nature_2021}.

\begin{equation}
f_\text{sync}\approx f_1 + \frac{\kappa_1}{\kappa_1+\kappa_2}(f_2-f_1),
\end{equation}

where $\kappa_i$ are effective coupling weights. Thus, the synchronized frequency interpolates between the two natural frequencies, leaning toward the one with stronger coupling, higher amplitude, or lower damping.

\item \textbf{Frequency pulling:} the two peaks move toward each other and broaden. Partially seen in see Fig.~\ref{figSI2}b for the fundamental frequencies and higher harmonics.

\item \textbf{Free-running:}
two distinct peaks at $f_1$ and $f_2$; in a combined signal you may also see intermodulation products $(n f_1 \pm m f_2)$ and a time-domain beat at $(|f_1 - f_2|)$, see Fig.~\ref{figSI2}c, Fig.~\ref{figSX}, and Fig.~\ref{figSY}a.

\end{itemize}

Figure~\ref{figSI2}a shows the extended frequency range, demonstrating how the synchronization is seen for the first three harmonics of the signal. The green colored trace demonstrates peaks in the middle between the pump-1-only and pump-2-only harmonics.

Figure~\ref{figSI2}b shows the extended frequency range in the case where the synchronization is broken. The green-colored trace demonstrates a double-peaked structure for each harmonic, with one peak corresponding to the frequency of pump-1 and the second peak being close to the pump-2-only position (although slightly shifted).

Figure~\ref{figSI2}c is the expanded view of Fig.~\ref{figSI2}b for the case without synchronization with vertical lines marking the main harmonics, the beat, and intermodulation products.

Figure~\ref{figSX} demonstrates the FFT peaks positions for all cases of two-pump beam separations, presented in Fig.~\ref{fig3}f of the main text. The presence of a single peak in the both-pumps-on case indicates synchronization, although the peak position does not necessarily coincide with the exact average of the individual frequencies. In the absence of synchronization, two distinct peaks corresponding to the contributing frequencies, along with their intermodulations, can be observed.

\section{Interaction by cross-illumination}

A pump light scattered from one spot could reach the other spot and provide a (very) weak common drive. This regime cannot be distinguished from the others in an experiment with a single pump, as shown in Fig.\,2, where different ENSSs are excited using a single pump spot. Still, we can be certain of a clear separation between the spots in experiments with two pumps, as shown in Fig.\,3, excluding this mechanism.

\begin{figure*}[h]
\begin{center}
\includegraphics[width=14cm]{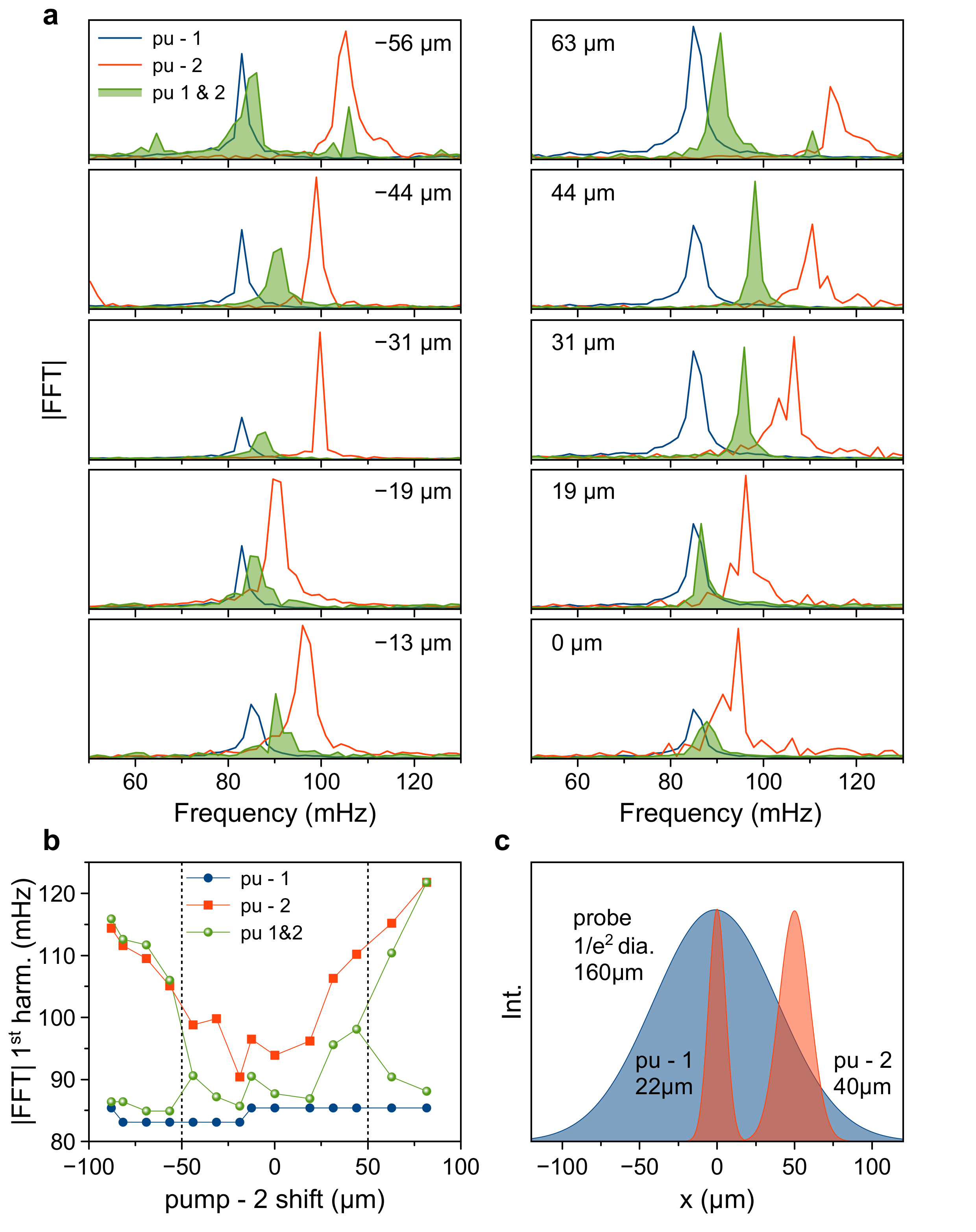}
\caption{\label{figSY} \textbf{Synchronization with spatially wide beams.}
\textbf{a}, FFT spectra for pump-1-only (blue), pump-2-only (red), and both pumps together (green shaded) cases for different pump-spot separations.
\textbf{b}, A summarizing plot with points representing the central frequency of the first FFT harmonic. Vertical dashed lines at $\pm 50\,\mu$m indicate the approximate synchronization border.
\textbf{c}, Simulation of the measured spot sizes with the probe beam diameter 160\,$\mu$m at $1/e^2$ intensity roll off, blue shaded region. Pump-1 is placed at the center of the probe and has a diameter of 22\,$\mu$m, while pump-2 is shifted by 50\,$\mu$m away from pump-1 and has a diameter of 40\,$\mu$m.
}
\end{center}
\end{figure*}

To demonstrate the influence of the cross-illumination, we have expanded both pump spots as shown in Fig.~\ref{figSY}c. Figure~\ref{figSY}a demonstrates exemplary FFT spectra at different pump-separations, while Fig.~\ref{figSY}b summarizes the measurements. In that setting, the observed synchronization radius between the pump spot centers has increased to about $50\pm5\,\mu$m. Taking the spot sizes into account, this leads to a beam separation of $19\pm5\,\mu$m, defined as the distance between the points where the spot intensities fall to $1/e^2$ of their maximum values. This value is close to that observed for the spatially narrow pump spots of $25\pm 3\,\mu$m, and gives additional support for our observations.

\section{Spin diffusion calculations}

\begin{figure*}[h]
\begin{center}
\includegraphics[width=14cm]{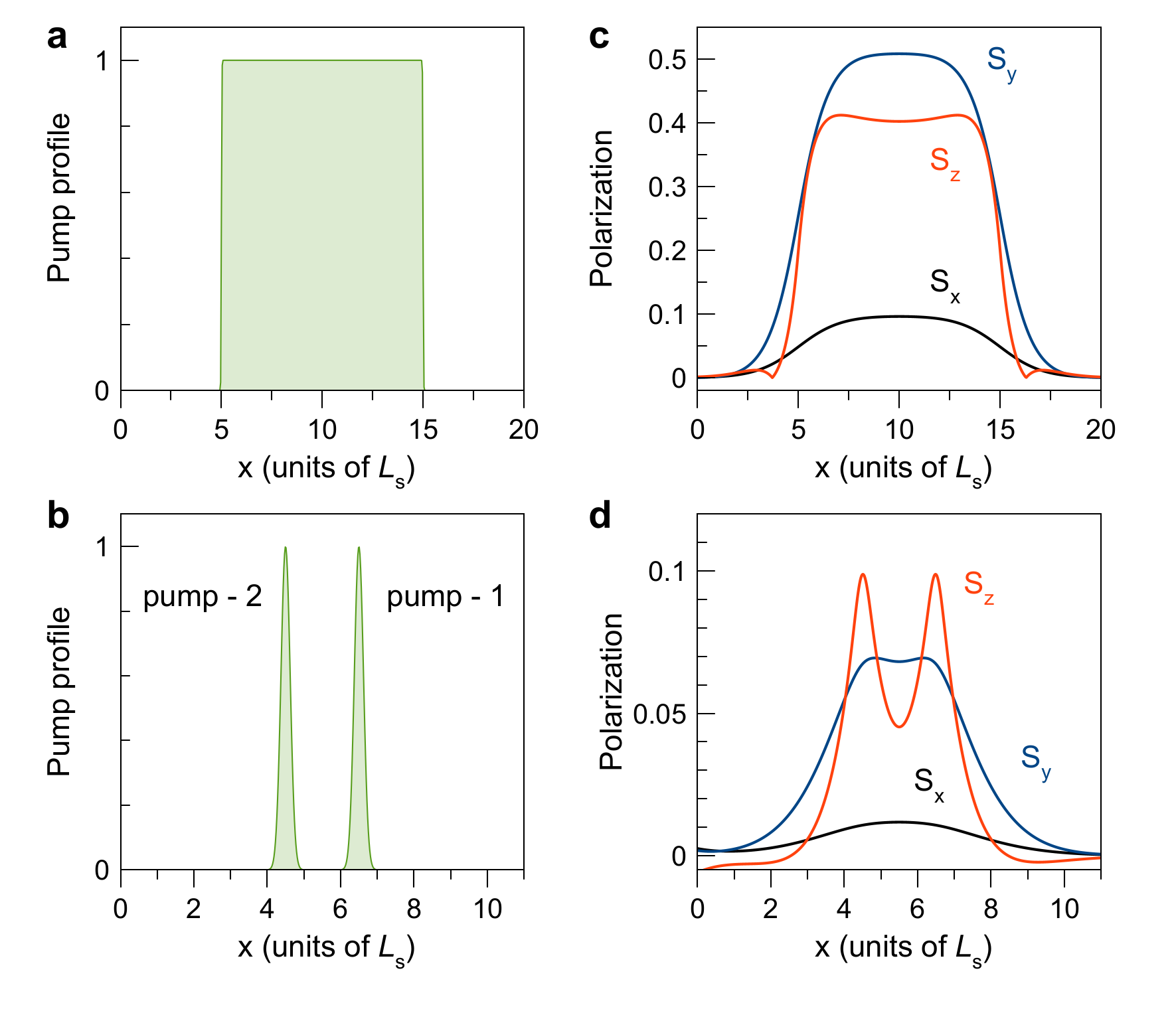}
\caption{\label{figSIT1} \textbf{Electron spin diffusion in magnetic field.} Spatial profiles of the flat-top pump covering the range from $x=0$ to $x=10L_\text{s}$ (\textbf{a}) and two Gaussian pumps having FWHM of $0.25L_\text{s}$ (\textbf{b}) separated by a distance of $2L_\text{s}$. Electron spin polarization components: $S_\text{x}$ (black), $S_\text{y}$ (blue), and $S_\text{z}$ (red), in a magnetic field of $B_x = -1$\,mT ($\alpha = 10^\circ$) under excitation by the flat-top pump (\textbf{c}) and by two Gaussian pumps (\textbf{d}).}
\end{center}
\end{figure*}

Figure~\ref{figSIT1}a shows the spatial profile of the flat-top pump covering the region from $x = 0$ to $x = 10L_\text{s}$ used for calculations of Fig.~4f in main text. The corresponding components of the electron spin polarization in the external magnetic field $B_x = -1$\,mT tilted by $\alpha = 10^\circ$ with respect to the sample plane, $S_\text{x}$ (black), $S_\text{y}$ (blue), and $S_\text{z}$ (red), are shown in Fig.~\ref{figSIT1}b.

For calculations of Figs.~4a-e, the excitation of two Gaussian pumps with a full width at half maximum of $0.25L_\text{s}$,
separated by a variable distance, is used. Figure~\ref{figSIT1}c illustrates their spatial profile. The components of the electron spin polarization, $S_\text{x}$ (black), $S_\text{y}$ (blue), and $S_\text{z}$ (red), are shown in Fig.~\ref{figSIT1}d for excitation by the two Gaussian pumps in the external magnetic field $B_x = -1$\,mT tilted by $\alpha = 10^\circ$ with respect to the sample plane.

\begin{figure*}[h]
\begin{center}
\includegraphics[width=17cm]{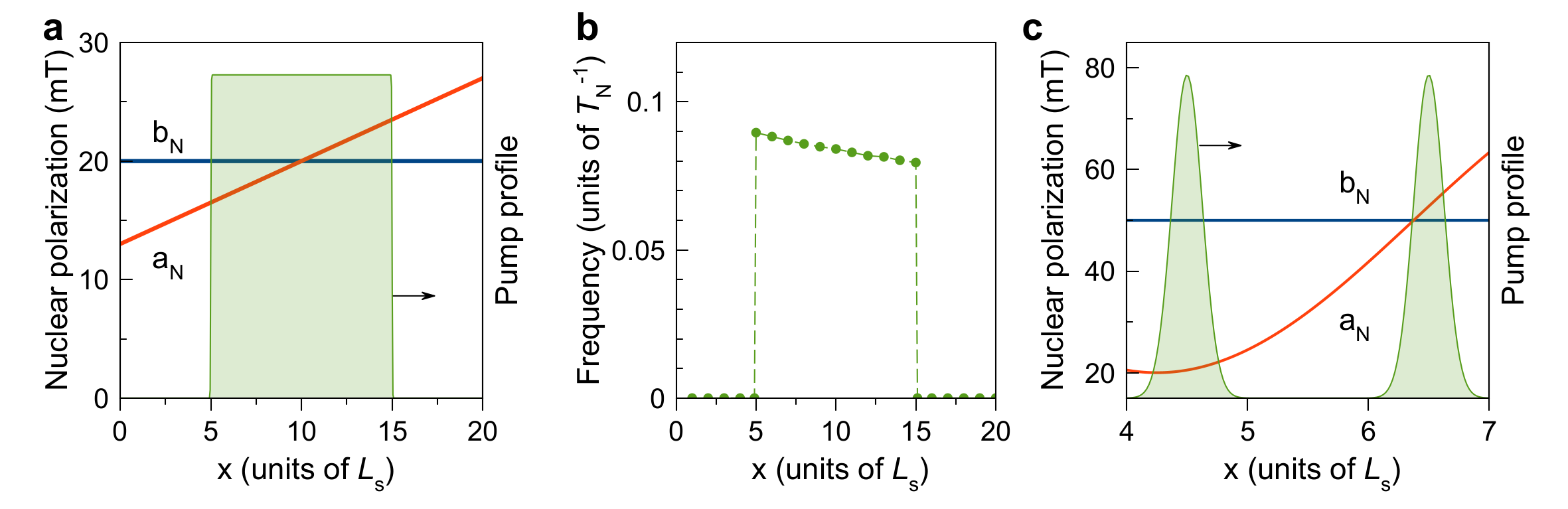}
\caption{\label{figSIT2}\textbf{Spatial distribution of hyperfine interaction parameters.}
\textbf{a}, Spatial dependence of $a_\text{N}$ (red) and $b_\text{N}$ (blue) under flat-top pump excitation; the green curve indicates the pump profile.
\textbf{b}, Spatial dependence of the first harmonic for hyperfine interaction parameters shown in \textbf{a} without spin diffusion.
\textbf{c}, Spatial dependence of $a_\text{N}$ (red) and $b_\text{N}$ (blue) under excitation by two Gaussian pumps separated by $2L_\text{s}$; the green curve indicates the pump profiles.
}
\end{center}
\end{figure*}

Figure~\ref{figSIT2}a shows the spatial dependence of $a_\text{N}$ (red) and $b_\text{N}$ (blue) under flat-top pump excitation (green), used for calculations of Fig.~4f in the main text. Figure~\ref{figSIT2}b presents the corresponding spatial dependence of the first harmonic calculated for the hyperfine interaction parameters shown in panel~\textbf{a} in the absence of spin diffusion in the external magnetic field $B_x = -1$\,mT tilted by $\alpha = 10^\circ$ with respect to the sample plane. Figure~\ref{figSIT2}c displays the spatial dependence of $a_\text{N}$ (red) and $b_\text{N}$ (blue) under excitation by two Gaussian pumps separated by a distance of $2L_\text{s}$, with the green curve indicating the respective pump profiles. The relative values of the hyperfine interaction parameters were increased compared to those used in the calculations with the flat-top pump. This adjustment was necessary because the spin polarization generated by a spatially narrow Gaussian pump is smaller than the average polarization produced by the spatially broad flat-top pump.
To ensure that the resulting frequency of the first harmonic in the Fourier spectrum remains comparable between the two calculation types, the hyperfine coupling constants were adjusted.

\begin{figure*}[h]
\begin{center}
\includegraphics[width=15cm]{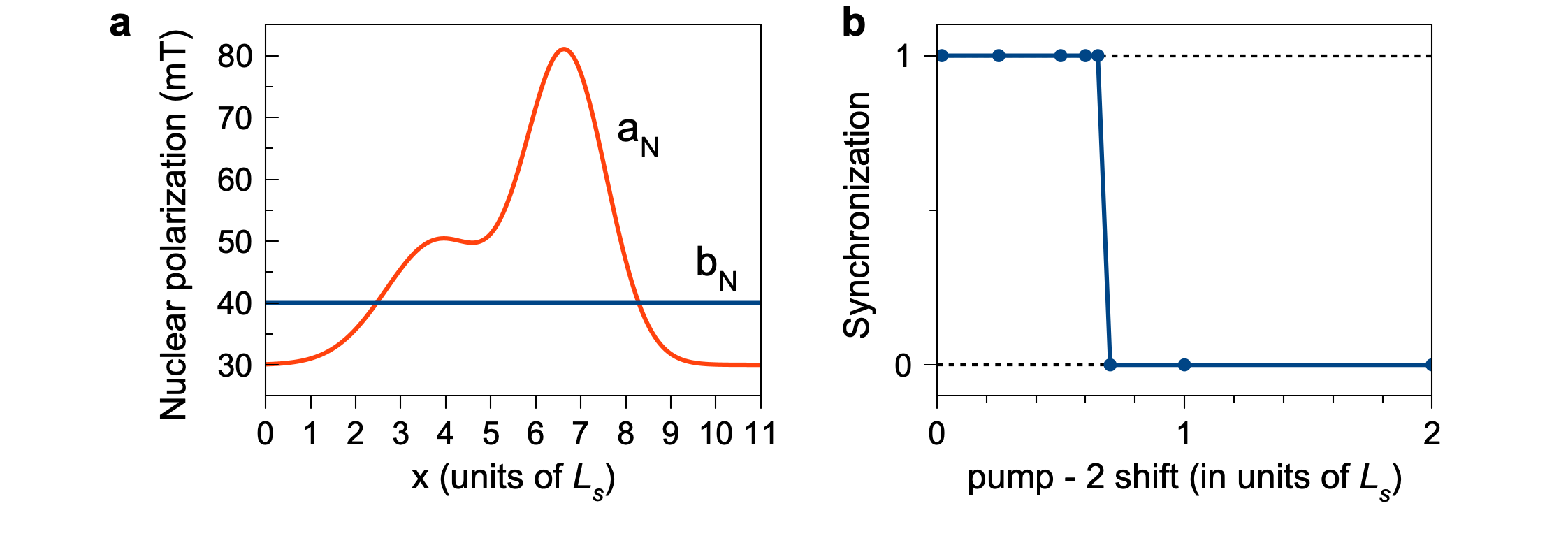}
\caption{\label{figSIT3}\textbf{Spatial distribution of hyperfine interaction parameters.}
\textbf{a}, Spatial dependence of $a_\text{N}$ (red) and $b_\text{N}$ (blue).
\textbf{b}, Synchronization range for $a_\text{N}$ and $b_\text{N}$ from panel {\bf a} for pump-2 shifts between 0 and 2$L_\text{s}$ relative to pump-1.
}
\end{center}
\end{figure*}

To highlight the dependence of the synchronization range on the hyperfine coupling span, we present an additional calculation. Figure~\ref{figSIT3} shows an example of the spatial distribution of hyperfine interaction parameters. The difference between the minimum and maximum values of \( a_N \) is now increased up to \( 50\,\text{mT} \), which is larger than in the distribution used for the calculations presented in the main text. In this case, the synchronization threshold shifts down to \( 0.65\,L_s \).

\section{Spatial Kerr microscopy}

\begin{figure*}[h]
\begin{center}
\includegraphics[width=16cm]{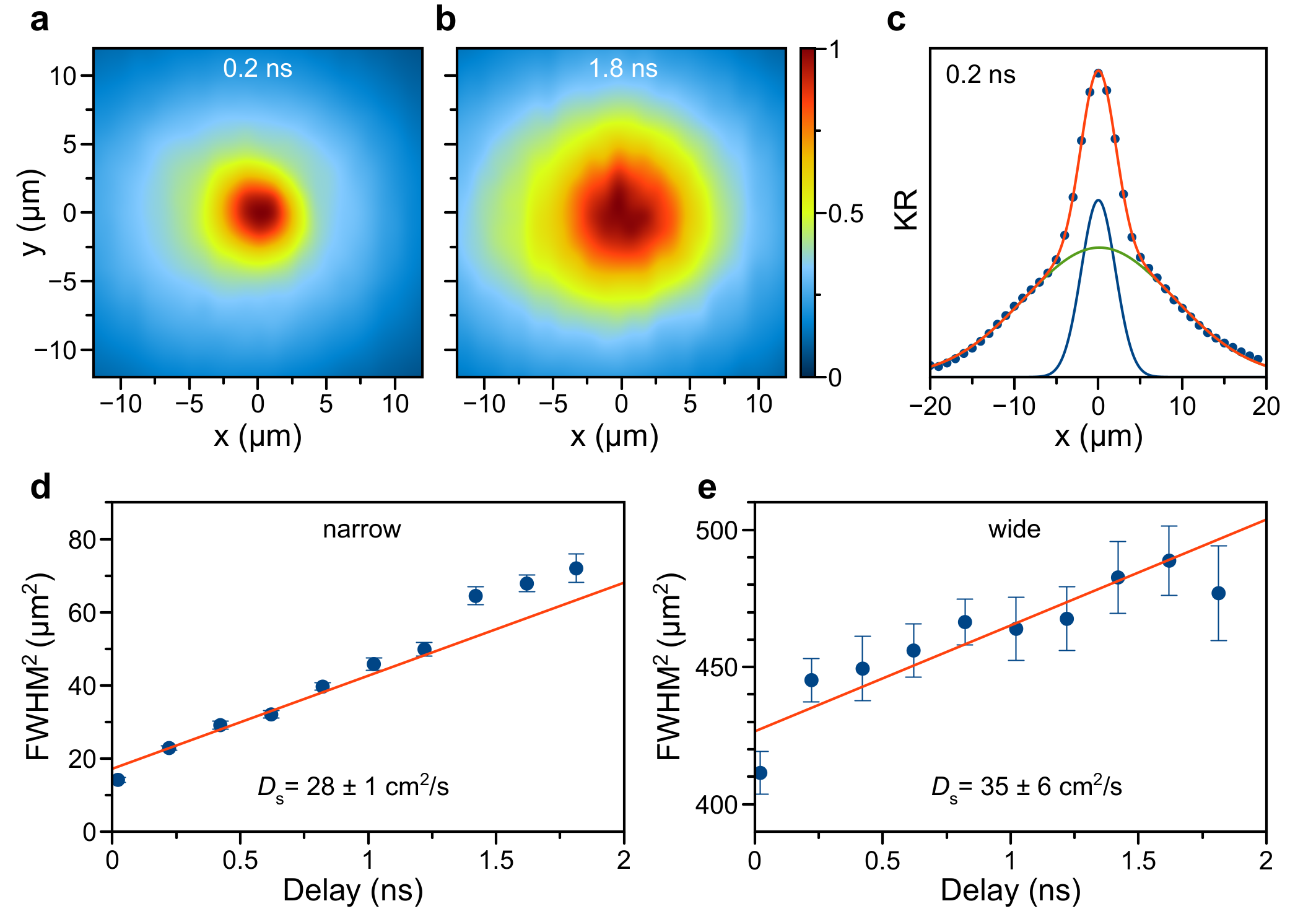}
\caption{\label{figSI4} \textbf{Full electron spin diffusion.}
Contour plots for the time-resolved pump-probe Kerr rotation experiment with spatial mapping measured at time-delays of 0.2\,ns~(\textbf{a}) and 1.8\,ns~(\textbf{b}), demonstrating the evolution of the spatial profile of the excited spin polarization. The Kerr rotation amplitude is normalized, and its scale is shown by a color bar on the right.
\textbf{c}, Exemplary double Gaussian fit of the spin polarization at the time delay of 0.2\,ns. The blue data points are well-fitted by a narrow (blue line) and a wide (green line) Gaussian profile. The red line is the sum of both fits.
\textbf{d}, Extracted time evolution of the narrow fit components in FWHM$^2$ versus time delay. The slope of the linear fit yields the spin diffusion coefficient $D_s=28\pm 1\,$cm$^2$/s.
\textbf{e}, The same for the wide fit component, with $D_s=35\pm 6\,$cm$^2$/s.
}
\end{center}
\end{figure*}

Figures~\ref{figSI4}a and~\ref{figSI4}b demonstrate the spatial maps of the time-evolution of the excited spin polarization measured by the pump-probe scanning Kerr microscopy at time delays of 0.2 and 1.8\,ns, exemplary. The images directly reveal the spatial dependence of spin diffusion after the excitation pulse of 1\,ps duration.

Figure~\ref{figSI4}c demonstrates the exemplary space-resolved spin polarization, derived from the measured Kerr rotation at 0.2\,ns time delay in the spatial Kerr microscopy. As shown, the space-resolved polarization consists of two Gaussian profiles: a narrow one is produced by the pump pulse at zero time delay and becomes broader with increasing time delay. In contrast, the wider profile is related to the signal coming from the previous-pump-pulse contributions, similar to Ref.~\cite{Henn_PRB2013}. We have fitted both of them separately. The corresponding dependencies of the FWHM$^2$ versus time delay are shown in Figs.~\ref{figSI4}d and~\ref{figSI4}e, yielding similar values of diffusion coefficients: $D_s=28\pm 1\,$cm$^2$/s and $D_s=35\pm 6\,$cm$^2$/s, respectively.

Furthermore, following the discussion in Ref.~\cite{Henn_PRB2013}, one can estimate the influence of hot electrons. As excitation occurs at much higher energies than the detection, the pump excites hot electrons that relax/cool down with an accelerated diffusion constant. According to Ref.~\cite{Henn_PRB2013}, one expects the electrons to cool down in approximately 400\,ps, while the diffusion constant is changing exponentially (decreasing) within that time. So, over the $\sim 400\,$ps cooling window, the enhanced diffusion increases the excited spin-packet width by $\approx 3\,\mu$m. The hot-electron phase provides a slight additional contribution to diffusion, but in the grand scheme, the longer, cooler diffusion is the dominant length scale.

In the experiments presented in the main text, the excitation energy was chosen to be close to the probe energy, allowing us to measure the diffusion constant of the cooled electron spins.

\end{document}